\documentstyle[prl,aps,twocolumn,amssymb]{revtex} 

\newcommand{\r}{{\bf r}}

\begin{document}
\draft
\title{Structure and Stability of Vortices in Dilute 
Bose-Einstein Condensates at Ultralow Temperatures}
\author{S. M. M. Virtanen, T. P. Simula, and M.~M.~Salomaa}
\address{Materials Physics Laboratory, 
Helsinki University of Technology\\
P.~O.~Box 2200 (Technical Physics), FIN-02015 HUT, Finland}
\date{\today}

\maketitle
\begin{abstract}
We compute the structure of a quantized vortex line in a 
harmonically trapped dilute atomic Bose-Einstein condensate using 
the Popov version of the Hartree-Fock-Bogoliubov 
mean-field theory. The vortex is shown to be (meta)stable 
in a nonrotating trap even in the zero-temperature limit,
thus confirming that weak particle interactions induce the condensed gas
a fundamental property characterizing ``classical'' superfluids. 
We present the structure of the vortex at ultralow temperatures 
and discuss the crucial effect of the thermal gas component to its 
energetic stability.
\end{abstract}
\hspace{5mm}

Dilute atomic Bose-Einstein condensates (BECs) have been the subject
of numerous experimental and theoretical studies since the first
landmark experiments in 1995 \cite{first_exp1,first_exp2,first_exp3}. 
These quantum fluids are unique in providing an opportunity to investigate 
how the phenomenon of Bose-Einstein condensation is affected by weak 
particle interactions. The fundamental question, whether superfluidity 
can be sustained by particle interactions in such systems, has still 
remained partly open despite extensive research. 

The existence of stable, dissipationless vortices is a feature 
characteristic to superfluid behavior. This aspect on the 
superfluid properties of dilute BECs is especially
timely due to the recent experimental realizations of vortices in
such systems \cite{vortex_exp1,vortex_exp2}. The stability 
of vortex structures has been under vigorous theoretical analysis
\cite{stringari1,dalf_stringari,IM_stab0,rokhsar,fetter_lt_stab,svid_fetter_stab,IM_stab1,IM_stab2,herra_puu,stringari2,butts_rokhsar,feder,garciat}. 
It has been shown that vortices in cylindrically trapped condensates 
are unstable within the Bogoliubov mean-field approximation 
\cite{IM_stab0,rokhsar,fetter_lt_stab,butts_rokhsar,dodd_eigenfr} 
unless the system is continuously driven by a suitable rotating 
perturbation \cite{fetter_lt_stab,IM_stab1,stringari2}. On the other
hand, by taking into account effects of the thermal gas component in
the system, vortices have been shown to become (meta)stable at high enough 
temperatures even in a nonrotating trap, or when a suitable
external pinning potential is applied \cite{IM_stab2}.

The absence of dissipation in the superfluid flow implies that a 
circulating current persists even when the system is not rotated
by an external perturbation. ``Classical'' superfluid behavior 
thus implies stability of vortices even in a nonrotating vessel.
Strictly speaking, such states are local minima of free energy and
only metastable, the global minimum naturally corresponding to a 
nonrotating state. However, in order to investigate the superfluid
properties of dilute boson condensates, it is just this kind of
local energetic stability of vortices that needs to be clarified.

In this Letter we present results of computations concerning 
the structure of a cylindrically trapped
dilute BEC containing a vortex line, and show that such states
are locally energetically 
stable at all temperatures $T<T_{\rm c}$ ($T_{\rm c}$ denotes
the critical temperature of condensation) even in a nonrotating trap
without additional pinning potentials. In this respect such systems 
are indeed shown to behave like ``classical'' superfluids. We also
discuss the obvious discrepancy between this result and the
predictions of the Bogoliubov approximation, and the role of
the thermal gas fraction in stabilizing the vortex state in
the zero-temperature limit.

We consider a dilute Bose-condensed gas consisting of atoms with mass $m$,
trapped by a radial harmonic potential 
$V_{\rm trap}(\r)=\frac{1}{2}m\omega_{r}^2r^2$ in cylindrical
coordinates $\r=(r,\theta,z)$. The particle interactions are modelled
by an effective low-temperature contact potential 
$V_{\rm int}(\r,\r')=g\delta(\r-\r')$, 
with the bare interaction constant $g$ and $s$-wave scattering length
for binary collisions $a$ related by $g=4\pi\hbar^2 a/m$.
Assuming the total particle number in the condensate to be large 
enough to justify grand
canonical formalism, the equilibrium condensate wavefunction
$\phi(\r)$ satisfies the generalized Gross-Pitaevskii (GP) equation
\cite{griffin_form}
\begin{equation}
\label{eq:GP}
\{{\mathcal H}_0+U_{\rm c}(\r)|\phi(\r)|^2+2U_{\rm e}(\r)\rho(\r)\}\phi(\r)
=\mu\phi(\r),
\end{equation}
where ${\mathcal H}_0=-\hbar^2\nabla^2/2m+V_{\rm trap}(\r)$ is
the bare single-particle Hamiltonian for the trap, $\mu$ the chemical
potential, and $\rho(\r)$ the density of the noncondensed gas.
Functions $U_{\rm c}(\r)$, $U_{\rm e}(\r)$ are effective interaction couplings
for various mean-field approximations---for the Popov version they are
simply chosen as $U_{\rm c}(\r)=U_{\rm e}(\r)\equiv g$. In addition,
we have performed computations within so-called G1 and G2 approximations,
which are gapless mean-field theories taking into account effects of
the background gas on colliding atoms, neglected in the Popov 
approximation \cite{gener1,gener2}.
In the G1 version $U_{\rm e}(\r)\equiv g$ and
$U_{\rm c}(\r)=g[1+\Delta(\r)/\phi^2(\r)]$, where $\Delta(\r)$ is
the anomalous average of two Bose field operators; for G2 one chooses 
$U_{\rm e}(\r)=U_{\rm c}(\r)=g[1+\Delta(\r)/\phi^2(\r)]$. 
The usual procedure of diagonalizing the mean-field Hamiltonian by 
a Bogoliubov transformation yields coupled eigenvalue equations for 
the bosonic quasiparticle amplitudes $u_q(\r)$, $v_q(\r)$, and 
the eigenenergies $E_q$ of the form \cite{griffin_form}
\begin{mathletters}
\label{eq:HFB}
\begin{eqnarray}
{\mathcal L}u_q(\r)+U_{\rm c}(\r)\phi^2(\r)v_q(\r) & = & E_q u_q(\r),\\
{\mathcal L}v_q(\r)+U_{\rm c}(\r){\phi^*}^2(\r)u_q(\r) & = & -E_q v_q(\r).
\end{eqnarray}
\end{mathletters}
Above, ${\mathcal L}\equiv {\mathcal H}_0 - \mu + 
2U_{\rm c}(\r)|\phi(\r)|^2+2U_{\rm e}(\r)\rho(\r)$ and
$q$ denotes quantum numbers specifying the quasiparticle states.
In addition, we have selfconsistency relations for the noncondensate
density $\rho(\r)$ and for the anomalous average $\Delta(\r)$:
\begin{eqnarray}
\rho(\r)&=&\sum_q [(|u_q(\r)|^2+|v_q(\r)|^2)n(E_q)+|v_q(\r)|^2],
\label{eq:self1}\\
\Delta(\r)&=&\sum_q [2u_q(\r)v_q^*(\r)n(E_q)+u_q(\r)v_q^*(\r)],
\label{eq:self2}
\end{eqnarray}
where $n(E_q)=(e^{E_q/k_{\rm B}T}-1)^{-1}$ is the Bose distribution function.
The anomalous average is ultraviolet divergent, and we renormalize it
by subtracting the last term in the sum of Eq.\ (\ref{eq:self2}) 
\cite{renorm}.

Considering a condensate penetrated by a single vortex line, we search
for solutions of the form $\phi(\r)=\phi(r)e^{im\theta}$, where $m$
denotes the number of circulation quanta of the vortex. In this Letter 
we restrict to the case $m=1$ due to instability of multiquantum vortices
\cite{herra_puu,butts_rokhsar,garciat}. By utilizing
the cylindrical symmetry, Eqs.\ (\ref{eq:GP}) and (\ref{eq:HFB}) can be reduced
to radial equations, which we discretize using a finite-difference method.
Dirichlet boundary conditions are imposed at $r=R$, the radius $R$ chosen
large enough for finite size effects for the structure of the vortex 
to be negligible. In the $z$-direction we impose periodic boundary conditions
at $z=\pm L/2$, thus modeling a system in the limit
of a very weak axial trapping potential. Due to cylindrical symmetry,
the quasiparticle amplitudes can be chosen to be of the form
\begin{mathletters}
\label{eq:ansatz}
\begin{eqnarray}
u_q(\r)&=&u_q(r)e^{iq_z(2\pi/L)z+i(q_{\theta}+m)\theta},\\
v_q(\r)&=&v_q(r)e^{iq_z(2\pi/L)z+i(q_{\theta}-m)\theta},
\end{eqnarray}
\end{mathletters}
where $q_{\theta}$ and $q_z$ are integer angular and axial 
momentum quantum numbers, respectively. Discretization transforms 
Eqs.\ (\ref{eq:HFB}) to a narrow-banded matrix eigenvalue problem, which we 
solve using the Lanczos algorithm implemented in the ARPACK subroutine 
libraries \cite{arpack,our_paper}.
The nonlinear Gross-Pitaevskii equation is solved using
finite-difference discretization and an overrelaxation method. 
For a given value of the chemical potential $\mu$, the solution of
the GP equation and the noncondensate density are integrated 
to find out the total particle number, and the process is iterated 
until the chemical potential corresponds to the preset total 
number of particles. The solution of the GP equation can be mapped to a 
zero-energy solution of Eqs.\ (\ref{eq:HFB}), thus providing a test for 
the accuracy and consistency of the numerical methods used to solve 
these equations. We search selfconsistent solutions for 
Eqs.\ (\ref{eq:GP})--(\ref{eq:self2}) using an iterative scheme: 
The condensate wavefunction and chemical potential corresponding 
to a preassigned total number of particles are computed by solving 
the GP equation. Using the quasiparticle states obtained
from Eqs.\ (\ref{eq:HFB}), new mean-field potentials are computed using the
selfconsistency equations. The whole procedure is repeated until 
convergence to a desired accuracy. 
In order to stabilize the iteration at low temperatures,
we use underrelaxation in updating the mean-field potentials. 

The physical parameter values for the gas and the trap were chosen to
be the same as in Ref.\ \cite{IM_stab2}. The gas consists of sodium
atoms with mass $m=3.81\times 10^{-26}$ kg and $s$-wave scattering
length $a=2.75$ nm. The frequency of the trap is 
$\nu_r=\omega_r/2\pi=200$ Hz, and the density of the system is
determined by treating $N=2\times 10^5$ atoms in the computational 
domain with dimensions $R=20$ $\mu$m and $L=10$ $\mu$m.
The critical condensation temperature for the system is approximately
$T_{\rm c}\approx 1$ $\mu$K. The spatial grid for discretizing
the quasiparticle eigenequations was chosen dense enough to
guarantee a relative accuracy of $10^{-3}$ for the energy
of the lowest excitation, and better than $10^{-4}$ for the
other states. 

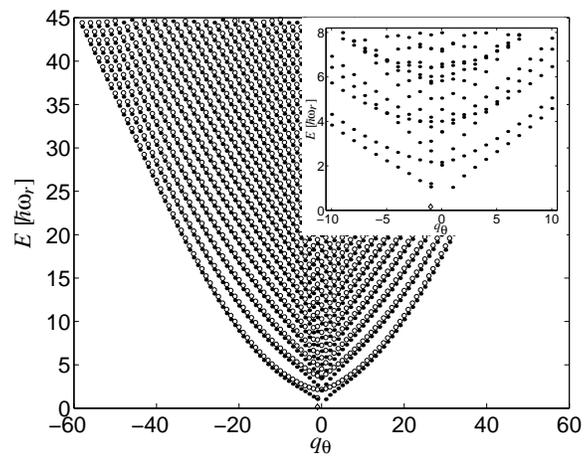
\begin{figure}
\caption{Part of the selfconsistent quasiparticle excitation spectrum
at temperature $T=100$ nK for the harmonically trapped condensate containing 
a single-quantum vortex line. Only states with axial momentum quantum
numbers $q_z=0$ (dots) and $q_z=1$ (circles) are shown.
The lowest Kelvin mode state, the LCLS, is denoted by a diamond.
The inset presents a blow-up of the lowest part of the full spectrum.
}
\label{fig:spectrum}
\end{figure}

Results of the selfconsistent computations performed
 within the Popov approximation
are shown in Figs.\ 1--4.
Figure \ref{fig:spectrum} displays part of the selfconsistent 
quasiparticle excitation spectrum for
the condensate vortex state at the temperature $T=100$ nK.
Only states with $q_z=0,1$, which contain the excitations of lowest energy,
are shown. These states determine the local energetic stability of the
vortex configuration: If there exists a quasiparticle excitation
with negative energy, the condensate can lower its energy by
exciting this negative bosonic mode, the state thus being prone to collapse.
For single-quantum vortices the lowest excitations with $q_{\theta}=-1$,
the so-called Kelvin mode states,
form standing waves localized at the vortex core, the energies of these
states being substantially lower than those for other angular 
momentum quantum numbers. The sign of the energy of the lowest Kelvin
mode excitation, the lowest core localized state (LCLS), determines
the local energetic stability of the vortex. The LCLS is denoted by 
a diamond in Fig.\ \ref{fig:spectrum}. Due to the 
very low energy of the LCLS compared to the other excitations, 
its contribution to the mean-field potentials in the core region 
becomes dominant in the zero-temperature limit. On the other hand, 
even minor changes in the potential functions affect the energy of 
the LCLS, thus altering its contribution drastically via the 
sensitive Bose factor. This behavior makes the iteration process 
extremely delicate at low temperatures \cite{IM_stab2}, requiring
the use of suitable underrelaxation methods.

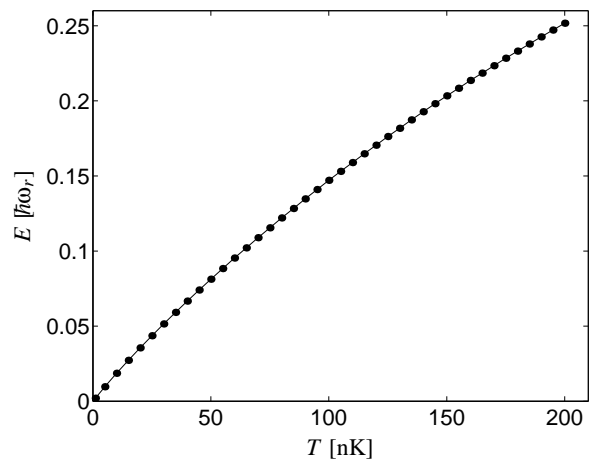
\begin{figure}
\caption{Energy of the lowest excitation, the LCLS, 
at temperatures $T=1$--$200$ nK ($T_{\rm c}\approx 1$ $\mu$K).
The finite positive value of the lowest excitation frequency implies 
the vortex to be locally energetically stable.
}
\label{fig:LCLS}
\end{figure}

The temperature dependence of the excitation energy of the LCLS 
is shown in Fig.\ \ref{fig:LCLS}. By carefully adjusting underrelaxation
in the iteration procedure, we are able to compute the selfconsistent
structure of the vortex down to the temperature of $T=1$ nK. 
The convergence at these temperatures is slow, but unquestioned. 
Especially, it was confirmed that the iteration converges to the 
same solution irrespective of the initial Ansatz used. However, 
it seems obvious that with additional computational effort one could reach
even lower temperatures. The result shows that although the energy
of the LCLS approaches zero in the low-temperature limit, as expected,
the vortex configuration is locally stable at all temperatures
$T<T_{\rm c}$ \cite{IM_stab2}. We obtained the same result also using
the G1 and G2 approximations.

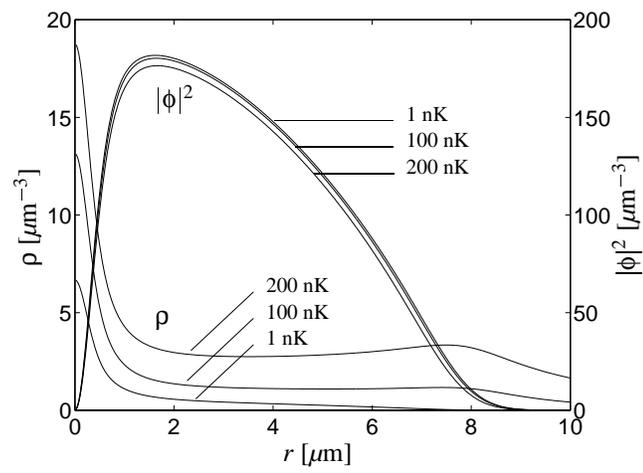
\begin{figure}
\caption{Condensate and thermal gas density profiles at temperatures
$T=1$ nK, $100$ nK, and $200$ nK. The noncondensate fills the space
left by the condensate at the vortex core, thus exerting an outward
pressure which stabilizes the vortex configuration.
}
\label{fig:structure}
\end{figure}

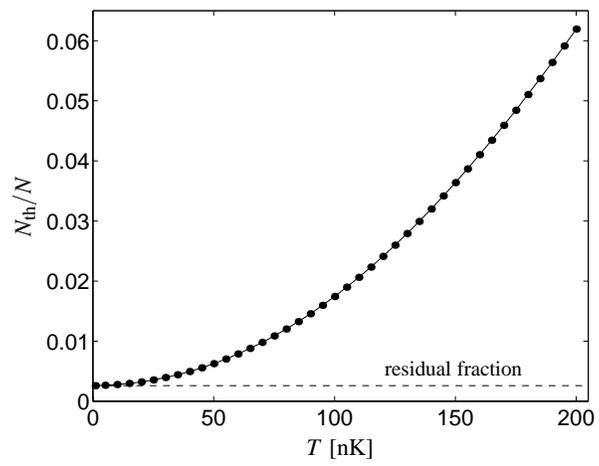
\begin{figure}
\caption{Noncondensate fraction as a function of
temperature for the vortex configuration. Due to particle interactions,
there exists a residual thermal gas fraction even in the zero-temperature
limit.
}
\label{fig:noncfrac}
\end{figure}

The local energetic 
stability of the vortex within the Popov, G1 and G2 \cite{HFB}
approximations is to be compared with the instability predicted by 
the Bogoliubov approximation \cite{rokhsar}, which does not treat the
thermal part of the gas selfconsistently. Figure \ref{fig:structure} 
shows the density profiles of the condensate and the noncondensate 
at ultralow temperatures. The thermal gas is concentrated 
in the vortex core, where it fills the space left by the 
condensate and exerts an outward pressure on it, 
preventing the condensate from collapsing into the core. Selfconsistent
treatment of the thermal gas fraction is thus crucial in determining
the stability of vortices in such systems, as seen in the discrepancy
between the predictions of the Popov, G1 and G2, and, on the other hand,
the Bogoliubov approximation. At temperatures of the order of $T_{\rm c}$,
the validity of the Bogoliubov approximation is expected to be 
questionable due to the substantial thermal gas fraction. It is to
be noted that it can fail also in the low-temperature limit in
certain respects, the stability of the vortex state being one
example. This is due to the nonvanishing, residual noncondensate
fraction present in interacting systems even in the zero-temperature limit. 
Figure \ref{fig:noncfrac} displays the computed thermal gas fraction 
at ultralow temperatures. It clearly shows the residual thermal 
fraction, which stabilizes the vortex at ultralow temperatures.

It is to be noted that the positivity of the LCLS energy implies that
vortices slightly displaced from the symmetry axis of the trap precess
in the direction opposite to the condensate flow around the core 
\cite{prec_lcls}. 
Recent experiments, however, show precession in the direction of the
flow, except for a minority of the so-called rogue vortices \cite{new_exp}. 
This seems to imply a negative LCLS energy, in agreement with the 
zero-temperature Bogoliubov approximation. The apparent 
discrepancy between the experiments and the predictions of the 
selfconsistent approximations could be due to insufficient thermalization 
in the experiments of the gas in the region of the (moving) vortex core. 
We expect that investigation of the validity of this assumption will 
further clarify the role of quasiparticles in stabilizing vortices in
weakly interacting Bose-Einstein condensates.

In conclusion, we have computed within selfconsistent mean-field theories
the structure of a cylindrically trapped, dilute atomic Bose-Einstein
condensate penetrated by a vortex line. The vortex state is shown
to be locally energetically stable in a nonrotating system even
in the zero-temperature limit, thus confirming the system to act
in this respect like a superfluid. The thermal gas concentrated
in the vortex core is shown to have a crucial effect in stabilizing
the vortex state even at ultralow temperatures, due to a 
residual noncondensate fraction present in interacting systems in the 
zero-temperature limit.

We thank the Center for Scientific Computing for computer resources and
the Academy of Finland for support.

\newpage

\begin{references}
\bibitem{HFB}
The vortex state is known to be locally stable also in the 
Hartree-Fock-Bogoliubov theory \cite{IM_stab0}. This approximation
is, however, plagued by an unphysical gap in the excitation 
spectrum \cite{griffin_form,gener1,gener2}, which also compromises
the validity of its prediction for stable vortex structures.
\end{references}
\end{document}